\input  phyzzx
\vsize=9.0truein
\voffset=-0.1truein
\hoffset=-0.1truein

%
%

\def\IC{{\ \hbox{{\rm I}\kern-.6em\hbox{\bf C}}}}
\def\IR{{\hbox{{\rm I}\kern-.2em\hbox{\rm R}}}}
\def\IZ{{\hbox{{\rm Z}\kern-.4em\hbox{\rm Z}}}}

\def\sIR{{\hbox{{\sevenrm I}\kern-.2em\hbox{\sevenrm R}}}}

\def\sym{super Yang Mills theory}
\def\mt{Matrix theory}
\def\sug{supergravity}

\def\sbh{schwarzschild black hole}

\def\g{g_{ym}}
\def\l{l_{st}}
\def\s{\Sigma}
\def\el{l_{11}}
%
%
\hyphenation{Min-kow-ski}
\rightline{SU-ITP-98-??}
\rightline{hep-th/9805115}
\rightline{May 1997}

\vfill

%
%
\title{Matrix Theory Black Holes and the Gross Witten Transition}

\vfill

%
%
\author{L. Susskind$^1$}

\vfill

\address{$^1$Department of Physics,
Stanford University\break Stanford, CA
94305-4060}
\vfill


\vfill

\vfill

%
%
Large N gauge theories  have  so called Gross-Witten
phase transitions which typically can occur in finite volume systems. In this
paper we relate these transitions in supersymmetric  gauge theories to
transitions that take place between black hole solutions in general relativity.
The correspondence between gauge theory and gravitation  is through  matrix
theory which represents the gravitational system in terms of \sym \ on finite
tori. We also discuss a related transition that was found by Banks, Fischler,
Klebanov and Susskind.

\vfill\endpage

%
%
\REF\bfks{T. Banks,W. Fischler,I. Klebanov and L. Susskind, hep-th/9709091 .}
\REF\bfss{T. Banks,W. Fischler,S. Shenker and L. Susskind,  hep-th/9610043 .}

\REF\gw{D.J. Gross, E. Witten,Phys.Rev.D21:446-453,1980}

\REF\witt{ Edward Witten,    hep-th/9802150}

\REF\ks{Igor R. Klebanov, Leonard Susskind, Schwarzschild Black Holes in Various
Dimensions from Matrix Theory,  hep-th/9709108 }

\REF\emil{Gary T. Horowitz, Emil J. Martinec, Comments on Black Holes in Matrix
Theory,  hep-th/9710217 }

\REF\corres{Gary T. Horowitz, Joseph Polchinski, A Correspondence Principle for
Black Holes and Strings,  hep-th/9612146}

\REF\gavin{G. Polhemus, Statistical Mechanics of Multiply Wound D-Branes,
hep-th/9612130}


%
%

%
%
\chapter{The Gross Witten Transition}

According to matrix theory [\bfss], there is a duality between Super Yang
Mills theory on a spatial d-torus
and 11 dimensional supergravity compactified on a d+1 torus \footnote*{ We
include the longitudinal 11th direction among the compact directions. Going to
large N effectively decompactifies this direction but this is not the main
focus of this paper}. Our knowledge of the two theories sometimes overlaps and
this allows us to test the duality. More often, we know something about one
theory which leads to predictions about the other. In this paper we will use
knowledge about black holes to gain information about Gross Witten type
transitions [\gw] in 3+1 dimensional \sym \ compactified on a 3-torus of
size
$\Sigma$.  In particular we will see that such a transition exists and that the
transition temperature  is a function of the 't Hooft coupling
$g_{ym}^2 N$ which scales like
$$
T_{\gw} \sim  {1 \over \Sigma \sqrt{g_{ym}^2 N}}
\eqn\one
$$
for large values of $g_{ym}^2 N$. Detailed features of the transition could in
principle be predicted from classical solutions of \sug . A similar
argument has been given by Witten for the case of compactification
on a sphere. In this case the duality is between \sym \ and supergravity in
$AdS_5
\times S_5$ [\witt].

We begin with a brief review of Gross Witten transitions. Normally systems of
 small numbers of degrees of freedom or systems in finite
volume do not exhibit sharp phase transitions or singularities in the partition
function. However in the large $N$ limit such transitions become possible. In
fact the Gross Witten transition was
first found in the theory of a single plaquette in lattice gauge theory [\gw].

The plaquette is described by a unitary $N \times N$ matrix $U$. The energy is
$$
E= - {1 \over g^2} \tr U
\eqn\two
$$
and the partition function is
$$
Z= \int dU \exp \left[{{\beta \over g^2} \tr U} \right]
\eqn\three
$$

The coupling $g$ is assumed to satisfy 't Hooft scaling
$$
g^2 N = \lambda
\eqn\four
$$
with $\lambda$ fixed.
The integral over unitary matrices can be replaced by an integral over the $N$
eigenvalues $ \alpha_i = e^{i \theta_i} $ of $U$. The measure for the
integration is a certain determinant  $ \Delta $ whose essential properties
we will
describe. The partition function is
$$
Z= \int d \alpha \exp  \left[{N \beta \over \lambda} \sum_i \alpha_i + \log
\Delta \right]
\eqn\five
$$
The log of $ \Delta $ is a sum over pairs of $\alpha's$ of  a two body repulsive
potential which diverges logarithmically when the two eigenvalues approach one
another.
$$
log \Delta \sim  \sum  \log |\theta_i - \theta_j|
\eqn\six
$$

 The first term in the brackets in eq \four may be thought of as an
attraction of the eigenvalues to the point $\alpha = 1$. The equilibrium
behavior can be estimated by looking for a stationary point of the action which
occurs when the forces on an eigenvalue balance. The result at  is a ``droplet"
of eigenvalues with a density of order $N$. The size of the droplet depends only
on  $\beta \over\lambda$ and it has sharp edges as $N \to \infty$. For small
$\lambda$ the droplet is small and as $\lambda$ increases its size
increases until at some critical value of $\beta \over\lambda$ the edges of the
distribution meet at
$
\theta = \pi$. At this point there is a singularity. This is the Gross Witten
transition. For this very simple case the temperature of the transition scales
like
$$
T_c \sim {1 \over \lambda} = {1 \over g^2 N}
\eqn\seven
$$

In eq \seven \ we see the characteristic dependence of the transition
temperature on the 't Hooft coupling $g^2 N$ which serves as the expansion
parameter for perturbation theory in the large $N$ limit.

The description of the low temperature behavior as a droplet is a bit
misleading.
In fact the combination of repulsive short range forces and attraction to the
origin creates a lattice with regularly spaced eigenvalues that could be
described as a crystal. In fact even when the distribution spreads over the
circle at high temperature, a strong degree of local crystalline order is
preserved due
to the repulsive forces.

In this paper we will present evidence that Gross Witten transitions occur
in 3+1
dimensional  toroidally compactified super Yang Mills theory. In this case
the matrix
valued variables that replace the plaquette variable $U$ are the Wilson
loops around the
cycles of the torus.

%
%
\chapter{Matrix Black Holes}

In [\bfks]  matrix theory [\bfss]   was applied to the behavior of black holes
in M Theory compactified on a 3-torus of size $L$. Strictly speaking the theory
is compactified on a 4 torus but by passing to the large $N$ limit while keeping
$L$ fixed, the 11th direction is effectively decompactified. For the problem of
black holes the decompactification can be quantified as follows. Begin with a
configuration with entropy $S$. Here we assume that $S>>1$. For $N<S$ the
configuration described in [\bfks] behaves like a 10 dimensional near extremal
black hole with D0-brane charge. Alternatively it can be thought of as an 11
dimensional black string wound around the 11th direction. It is homogeneous in
the 11th direction.  As N increases, an instability is encountered. The black
string breaks and forms a localized blob in the 11th direction [\emil]. The blob
becomes an 11 dimensional Schwarzschild black hole. The transition from
homogeneous black string to localized black hole occurs at $N\sim S$. In fact
one finds that at this point, the free energy of the black string and
Schwarzschild black hole are equal.   We will return to this
transition in sect 3 but for now we are interested in the
"black string" region
$N<<S$ where the black hole is uniform in $x^{11}$.

Let us consider the behavior of the black hole as we vary the size of the
3-torus $L$, keeping $N$ large but fixed. It is convenient to think of
compactification
as periodic identification and to imagine a transverse lattice of black holes
with spacing
$L$. It is obvious that for very large
$L$ the compactification can be ignored and the black hole is localized
somewhere in the torus with a radius much smaller than $L$. In this region the
black hole behaves as a 10 dimensional near extremal D0-brane black hole in
uncompactified transverse space. It is so far from its periodic images that they
can be ignored. On the other hand for small enough
$L$ the black hole will merge with its images and a uniform homogeneous
configuration on the 3 torus will result. In fact a sharp transition must take
place [\corres]. To see this consider the horizon. For very large $L$ the
horizon
will form a small sphere localized in the torus. As $L$ is decreased the
horizons of the images will eventually touch and the black holes will merge. The
change in topology of the horizon is discrete and signals a singularity in its
area and thus in the thermodynamic quantities.

The approximate location of the phase transition can be found either by
thermodynamic or by solving the gravitational equations and determining the
point at
which the horizon changes topology. We will use thermodynamics. The
thermodynamic
relation between free energy and temperature for a near extremal D0-brane
black hole
is \footnote*{Throughout this paper all irrelevant numerical constants will be
set to 1}
$$
F={T^4 N^2 l_{st}^6 \over L^3}
\eqn\twoone
$$
where  $\l$ is the string length scale.

For a D0-brane black hole in uncompactified  ($L \to \infty $) space one finds
[\ks]
$$
F= T^{14/5} N^{7/5} l_{st}^{9/5}g_{st}^{-3/5}
\eqn\twotwo
$$
where $g_{st}$ is the string coupling constant.

To find the transition we equate \twoone \ and \twotwo \ to give
$$
T_{c}= \sqrt{L^3 \over \l^3 g_{st} N}
\eqn\twothree
$$

%
%
\chapter{Gauge Theory Interpretation of the Transition}

According to \mt \ the system we are describing is dual to \sym \
compactified on
a  3-torus. The parameters of the \sym \  are a coupling constant $\g$ and a
compactification radius $\s$. We also introduce two M-Theory quantities; $R$ and
$\el$. $R$ is the compactification radius of the 11 direction and  $\el$ is the
11 dimensional planck length. The various quantities are related by
$$\eqalign{
R&=\l g_{st} \cr
\el&=\l g_{st}^{1/3} \cr
}
\eqn\threeone
$$
and
$$
\eqalign{
\g^2&= {\el^3 \over L^3} \cr
\s& = {\el^3 \over L R}  \cr
}
\eqn\threetwo
$$
Plugging these equations into \twothree \ we find the transition temperature
satisfies
$$
\s T_c=(\g^2 N)^{-1/2}
\eqn\threethree
$$

Note that  the right side of eq \threethree \  depends only on the 't Hooft
coupling constant $\g^2 N$ as we would expect for a Gross Witten type
transition.
On the left side we the  scale invariant function of the temperature that we
can construct from the Yang Mills parameters. This reflects the exact conformal
invariance of  3+1 dimensional \sym \ with 16 real supersymmetries.

In order to see that this transition is analogous to the Gross Witten
transition on a single plaquette, let us recall the description of the black
hole (for $N<<S$) given in [\bfks ]. It was found that the D0-branes form a
regular lattice on the 3-torus. Indeed the 't Hooft parameter $\g^2 N = {N
\el^3 \over L^3}$ is the density of D0-branes. The origin of the lattice like
structure in both the 1-plaquette example and the much more sophisticated \sym \
is the same; eigenvalue repulsion [\gavin].

In interpreting the transition as a Gross Witten transition the most important
point is that the location of the D0-branes on the torus of size $L$ is given in
\mt \ by the eigenvalues of the Wilson loops around the 3 cycles. Thus, just as
for the single plaquette, the transition  is one in which the distribution of
Wilson loop eigenvalues begins at high temperatures by filling the torus with a
crystal lattice and passes  to a localized blob at low temperatures.

However the picture described above only holds over part of the $N,L$ plane. As
the temperature  is lowered it is possible that another transition intervenes
before the Gross  Witten  transition. This is the transition reported in [\bfks]
in which the 10 dimensional black hole (11 D black string) condenses into a
blob in the 11 direction and becomes an 11 dimensional \sbh. For want of a
better name the transition will be referred to as the BFKS transition. The BFKS
transition occurs at the point where the entropy is equal to $N$. The critical
temperature for this transition satisfies
$$
\s T_{bfks} = N^{-{1 \over 3}}
\eqn\threefive
$$
The gross Witten transition will occur only if $T_c> T_{bfks}$ or
$$
{L^9 \over \el^9}>N
\eqn\threefive
$$

Like the GW transition, the BFKS transition  also separates a homogeneous phase
from a blob-like phase.  At high temperature the black hole system will not only
be homogeneous in the 3-torus of size $L$ but also the longitudinal compact
direction of size
$R$. If the inequality \threefive \ is violated then as the temperature is
lowered the black object will become localized in the longitudinal direction.
This transition is not of the Gross Witten type in the 3+1 gauge theory but as
shown in [\bfks] its existence is easily understood in the D3-brane description
of the \sym \ as a transition which occurs when the thermal wavelength exceeds
the size of the effective quantization volume. For a more complete explanation
the reader is referred to [\bfks]. However in the  SYM description it is not
clear that this transition is a sharp one. The gravitational description
however,
makes it clear that a singularity occurs. The point is once again that when the
system goes from black string to black hole, the horizon topology suddenly
changes. This sudden change signals a singularity in the area and therefore the
entropy.

\refout
\end